\documentclass[preprint, 12pt]{elsarticle}

\usepackage[english]{babel}
\usepackage[utf8x]{inputenc}
\usepackage[T1]{fontenc}
\usepackage{amsthm}
\usepackage{paralist}
\usepackage{algorithm2e}
\usepackage{caption, subcaption}

 \AtBeginDocument{%
  \providecommand\BibTeX{{%
    \normalfont B\kern-0.5em{\scshape i\kern-0.25em b}\kern-0.8em\TeX}}}

\usepackage{amsmath}
\usepackage{graphicx}
\usepackage[colorinlistoftodos]{todonotes}
\usepackage{hyperref}
\usepackage{xparse}
\usepackage[super]{nth}
\usepackage{float}
\usepackage{multirow}
\usepackage{tikz}
\usepackage{graphicx}
\usetikzlibrary{arrows}
\usetikzlibrary{shapes}
\newcommand{\mymk}[1]{%
  \tikz[baseline=(char.base)]\node[anchor=south west, draw, circle, inner sep=1pt, minimum size=4mm, text height=2mm](char){\ensuremath{#1}};}

\usepackage{color}
\definecolor{editorGray}{rgb}{0.95, 0.95, 0.95}
\definecolor{editorOcher}{rgb}{1, 0.5, 0} %
\definecolor{editorGreen}{rgb}{0, 0.5, 0} %
\usepackage{upquote}
\usepackage{balance}
\usepackage{listings}
\lstdefinelanguage{JavaScript}{
  morekeywords={typeof, new, true, false, catch, function, return, null, catch, switch, var, if, in, while, do, else, case, break},
  morecomment=[s]{/*}{*/},
  morecomment=[l]//,
  morestring=[b]",
  morestring=[b]'
}

\lstdefinelanguage{HTML5}{
        language=html,
        sensitive=true, 
        alsoletter={<>=-},
        otherkeywords={
        <html>, <head>, <title>, </title>, <meta, />, </head>, <body>, <form, <input,
        <canvas, \/canvas>, <script>, </script>, </form>, </body>, </html>, <!, html>, <style>, </style>, ><
        },  
        ndkeywords={
        =,
        charset=, id=, width=, height=, method=, action=, type=, name=, value=, 
        border:, transform:, -moz-transform:, transition-duration:, transition-property:, transition-timing-function:
        },  
        morecomment=[s]{<!--}{-->},
        tag=[s]
}

\journal{Arxiv}
\begin{document}

\theoremstyle{definition}
\newtheorem{defn}{Definition}[section]

\theoremstyle{definition}
\newtheorem{rqn}{RQ}[section]

\theoremstyle{definition}
\newtheorem{note}{Note}[section]

\theoremstyle{remark}
\newtheorem{ex}{Example}[section]

\newcommand{\erratum}[0]{\textsc{Erratum}\xspace}
\newcommand{\erratumlong}[0]{\emph{rEpaiRing bRoken locATors Using tree Matching}}
\title{\erratum: \em Leveraging~Flexible~Tree~Matching to~Repair~Broken~Locators in~Web~Automation~Scripts}
\author{
    \emph{Sacha}~\textsc{Brisset}, 
    \emph{Romain}~\textsc{Rouvoy}, 
    \emph{Lionel}~\textsc{Seinturier}, 
    \emph{Renaud}~\textsc{Pawlak}
}

\begin{abstract}

Web applications are constantly evolving to integrate new features and fix reported bugs.
Even an imperceptible change can sometimes entail significant modifications of the \emph{Document Object Model} (DOM), which is the underlying model used by browsers to render all the elements included in a web application.
Scripts that interact with web applications (\emph{e.g.} web test scripts, crawlers, or robotic process automation) rely on this continuously evolving DOM which means they are often particularly fragile.
More precisely, the major cause of breakages observed in automation scripts are \emph{element locators}, which are identifiers used by automation scripts to navigate across the DOM. When the DOM evolves, these identifiers tend to break, thus causing the related scripts to no longer locate the intended target elements.

For this reason, several contributions explored the idea of automatically repairing broken locators on a page. 
These works attempt to repair a given broken locator by scanning all elements in the new DOM to find the most similar one.
Unfortunately, this approach fails to scale when the complexity of web pages grows, leading either to long computation times or incorrect element repairs.
This article, therefore, adopts a different perspective on this problem by introducing a new locator repair solution that leverages tree matching algorithms to relocate broken locators.
This solution, named \erratum{}, implements a holistic approach to reduce the element search space, which greatly eases the locator repair task and drastically improves repair accuracy.
We compare the robustness of \erratum{} on a large-scale benchmark composed of realistic and synthetic mutations applied to popular web applications currently deployed in production.
Our empirical results demonstrate that \erratum{} outperforms the accuracy of WATER, a state-of-the-art solution, by 67\%.
\end{abstract}

\begin{keyword}
Element~locator \sep Locator~repair \sep Robotic~process~automation \sep Tree~matching \sep Web~crawling \sep Web~testing
\end{keyword}

\maketitle

\section{Introduction}
The implementation of automated tasks on web applications (apps), like crawling or testing, often requires software engineers to locate specific elements in the DOM (\emph{Document Object Model}) of a web page.
To do so, software engineers or automation/testing tools often rely on CSS (\emph{Cascading Style Sheets}) or XPath selectors to query the target elements they need to interact with.
Unfortunately, such statically-defined locators tend to break along time and deployments of new versions of a web application.
This often results in the failure of all the associated automation scripts (including test cases) that apply to the modified web pages.

While several existing works focus on repairing tests on GUI applications, there are surprisingly very few test repair solutions targeting web interfaces~\cite{imtiaz2019systematic}.
These solutions either propose to \emph{i)} generate locators that are robust to changes (so-called \emph{robust locator problem}), or \emph{ii)} repair locators that are broken by the changes applied to the web pages (so-called \emph{locator repair problem}).
Unfortunately, most of the existing solutions in the literature fail to accurately relocate a broken locator, thus leaving all the related web automation scripts as broken~\cite{hammoudi2016record}.
More specifically, state-of-the-art solutions to the locator repair problem, WATER~\cite{choudhary2011water} and VISTA~\cite{stocco2018visual}, tend to rely on the intrinsic properties of the element whose locator needs repairing to locate its matching element on the modified page. 
However, this approach fails to leverage the element position and relations with the rest of the DOM, thus ignoring valuable contextual insights that may greatly help to repair the locator.

In this article, we adopt a more holistic approach to the locator repair problem: instead of focusing on the element whose locator is broken individually, we leverage a tree matching algorithm to match all elements between the two DOM versions. 
Intuitively, using a holistic approach to repair a broken locator should significantly improve accuracy by reducing the search space of candidate elements in the new version of the page: for example, if the parent of the element whose locator is broken is easily identifiable (\emph{e.g.}, the item of a menu) a tree matching algorithm will use this information to relocate the target element in the modified page with better accuracy.
Additionally, if more than one locator is broken on a given web page, our approach will repair all of them consistently at once.
The holistic solution we propose, named \erratum{},\footnote{\erratum{} stands for "\erratumlong{}"} more specifically leverages an efficient \emph{Similarity-based Flexible Tree Matching} (SFTM) algorithm we implemented to repair all broken locators by matching all changes in a web page with high accuracy.

Evaluating solutions to both robust locator and locator repair problems requires to build a dataset of web page versions---\emph{i.e.}, \textsf{(original page, modified page)} pairs.
Unfortunately, previous works assessed their contributions on hardly-reproducible benchmarks of limited sizes (never beyond a dozen of websites).
In this article, we rather evaluate the robustness of our approach against the state of the art by introducing an open benchmark, which covers a wider range of changes that can be found in modern web apps.
Concretely, our open benchmark considers over 83k+ locators on more than $650$ web apps.
It combines \emph{i)} a \emph{synthetic dataset} generated from random mutations applied to popular web apps and \emph{ii)} a \emph{realistic dataset} replaying real mutations observed in web apps from the Alexa~Top\,1K\footnote{\url{https://www.alexa.com/topsites}}, which ranks the most popular websites worldwide.

When evaluated on both datasets, our results demonstrate that \erratum{} outperforms the state-of-the-art solution, namely WATER~\cite{choudhary2011water}, both in accuracy (67\% improvement on average) and performances, when more than 3 locators require to be repaired in a web page.

Concerning the potential applications of \erratum, while we introduce and evaluate our solution within the well-studied context of locator repair, we also discuss a novel testing architecture centered around \erratum allowing to entirely replace all locator-based interactions.
This architecture intends to support much more interactive and robust script editions in the context of web testing, web crawling, and robotic process automation.

Overall, the key contributions of this article consist of:
\begin{compactenum}
    \item proposing a solution to the locator repair problem inspired from the \emph{Flexible Tree Matching} (FTM) algorithm,
    \item implementing and integrating an efficient extension of FTM algorithm to repair broken locators,
    \item providing a novel, reproducible, large-scale benchmark dataset to evaluate both the robust locator and locator repair problems,
    \item reporting on an empirical evaluation of our approach when solving the locator repair problem,
    \item proposing a novel script edition architecture centered on \erratum.
\end{compactenum}

The remainder of this article is organized as follows.
Section~\ref{sec:related-work} introduces the state-of-the-art approaches in the domain of robust locators and locator repair, before highlighting their shortcomings.
Section~\ref{sec:locator} formalizes the locator problem we address in this article.
Section~\ref{sec:implementation} introduces our approach, \erratum, which leverages an efficient flexible tree matching solution that we identified.
Section~\ref{sec:benchmark} describes the locator benchmark we designed and implemented.
Section~\ref{sec:evaluation} reports on the performance of our approach compared to the state-of-the-art algorithms.
Finally, Section~\ref{sec:perspectives} presents some perspectives for this work, while Section~\ref{sec:conclusion} concludes.

\section{Background \& Related Work}\label{sec:related-work}
We deliver a novel contribution to the locator repair problem, which has been initially studied in the domain of web testing.
In this section, we thus introduce the required background and we describe state-of-the-art approaches to repair broken locator, and in particular the literature published in the domain of web test repair.

\subsection{Introducing Web Element Locators}
To detect regressions in web applications, software engineers often rely on automated web-testing solutions to make sure that end-to-end user scenarios keep exhibiting the same behavior along changes applied to the system under test.
Such automated tests usually trigger interactions as sequences of actions applied on selected elements and followed by assertions on the updated state of the web page.
For example, \emph{"click on button $e_1$, and assert that the text block $e_2$ contains the text '{\tt Form sent}'"}.
To develop such test scenarios, a software engineer can
\begin{inparaenum}
    \item manually write web test scripts to interact with the application, or
    \item use record/replay tools~\cite{burg2013interactive,sen2013jalangi,mickens2010mugshot} to visually record their scenarios.
\end{inparaenum}
In both cases, the scenario requires to identify the target elements on the page [$e_1$, $e_2$] in a deterministic way, which is usually achieved using XPath, a query language for selecting elements from an XML document.
For example, let us consider the following HTML snippet describing a form:
\begin{lstlisting}
<form method="post" action="index.php"> 
    <input type="text"   name="username"/> 
    <input type="submit" value="send"/>
</form> 
\end{lstlisting}

The following XPath snippets describe 3 different queries, which all result in selecting the submit button: \lstinline!/form/input[2]!, \lstinline!/form/input[@value="Send"]!, \lstinline!input[@type="submit"]!.
In the literature, such element queries or identifiers are named \emph{locators}~\cite{leotta2014reducing}.

In practice, automated tests are often subject to breakages~\cite{hammoudi2016record}.
It is important to understand that a test \textit{breakage} is different from a test \textit{failure}~\cite{stocco2018visual}: a test failure successfully exposes a regression of the application, while a test is said to be broken whenever it can no longer apply to the application (\emph{e.g.}, the test triggered a click on a button $e$, but $e$ has been removed from the page).
While there can be many causes to test breakage, \cite{hammoudi2016record} reports that $74\,\%$ of web tests break because one of the included locators fails to locate an element in a web page.

\subsection{Generating Web Element Locators}\label{sec:sota_robus_locator}
The fragility of locators remains the root cause of test breakage, no matter they have been automatically generated (\emph{e.g.}, in the case of record/replay tools), or manually written.
To tackle this limitation, several studies have focused on generating more robust locators.
This includes ROBULA~\cite{leotta2014reducing}, ROBULA+~\cite{leotta2016robula+}, which are algorithms that apply successive refining transformations from a raw XPath query until it yields a locator that exclusively returns the desired element.
Leotta~\emph{et~al.}~\cite{leotta2015meta} also propose a graph-based algorithm to generate locators, but has not provided any evaluation to the algorithm.
Another work by Yandrapally~\emph{et~al.} leveraged contextual clues to generate locators~\cite{yandrapally2014robust}. These clues rely mostly on the content surrounding the element to locate which may be problematic in case the content changes.
LED~\cite{bajaj2015synthesizing} uses a SAT solver to select several elements at once, but is never evaluated on different DOM versions.
Finally, some works combine several locator generators with a voting mechanism to locate a single element with more robustness~\cite{leotta2015using, zheng2018method, long2020webrr}.
However, all these approaches, which consider a limited set of locator generators, strongly depend on the accuracy of individual algorithms to agree upon a single and relevant locator.
While automatically generating locators can speed up the definition of test cases, it becomes a keystone for visually-generated test cases based on record/replay tools.
In the end, the reliability of test cases built using such a tool depends mostly on the quality of the locators it automatically generates~\cite{hammoudi2016record}.

\subsection{Repairing Web Element Locators}
While some solutions to the robust locator problem, as presented above, aim to prevent locators from breaking, others focus on repairing broken locators.
In this context, the repair tool considers
\begin{inparaenum}[\em a)]
    \item the descriptor of the locator,
    \item the last version of the page on which the locator was still functional ($D$),
    \item the new version of the page on which the locator is broken ($D'$).
\end{inparaenum}

\paragraph{WATER}
In this area, WATER~\cite{choudhary2011water} provides an algorithm to fix broken tests.
The process of repairing a test involves several steps: 
\begin{inparaenum}
\item running the test, 
\item extracting the causes of failure and,
\item repairing the locator, if broken. 
\end{inparaenum}
The last part is particularly challenging. 
To relocate a locator from one version to another, WATER scans all elements in the new version and returns the most similar one to the element in the original version with regards to intrinsic properties (\emph{e.g.}, absolute XPath, classes, tag).
Hammoudi~\emph{et~al.}~\cite{hammoudi2016waterfall} further studied the locator repair part of WATER and found that repairing tests over finer-grained sequences of change (typically commits) contributes to improving accuracy.

\paragraph{VISTA}
Using a completely different approach, VISTA~\cite{stocco2018visual} is a recent technique that adopts computer vision to repair locators. VISTA falls within the category of computer vision-aided web tests~\cite{chang2010gui,leotta2018pesto,alegroth2013jautomate}.
However, while using computer vision succeeds in repairing most of the \textit{invisible} changes, such solutions tend to fail when the content, the language, or the visual rendering of the website changes.
Furthermore, visual-based solutions fail to locate dynamic elements that only appear through user interactions (\emph{e.g.}, a dropdown menu).

\begin{figure*}
    \centering
    \includegraphics[width=1\linewidth]{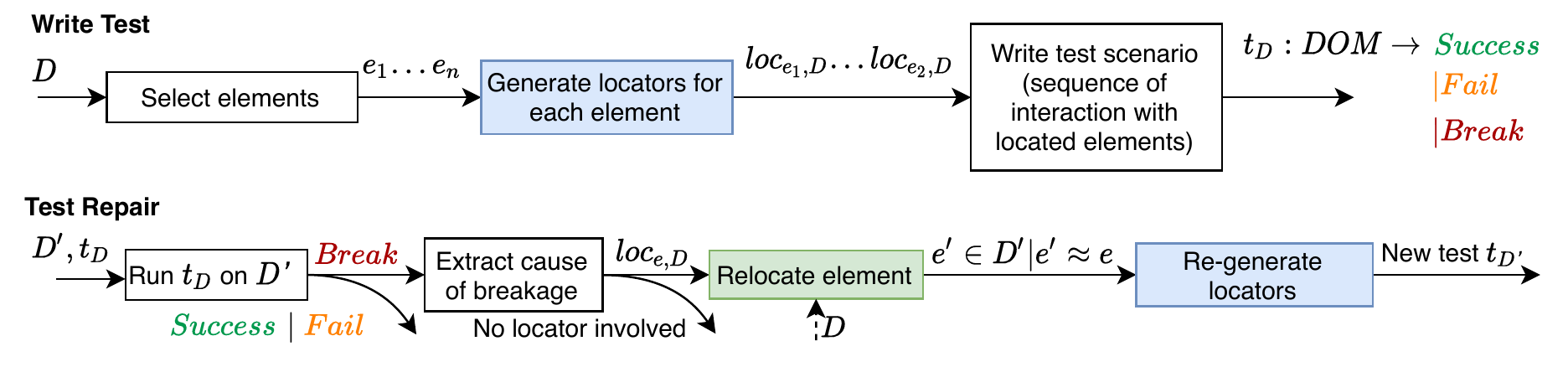}
    \caption{Illustration of the locator problem statement in automated tests combining the \emph{robust~locator}~(in~blue) and the \emph{locator~repair}~(in~green)~problems.}
    \label{fig:locator_repair}
\end{figure*} 

\section{Locator Problem Statement}\label{sec:locator}
Figure~\ref{fig:locator_repair} summarizes the steps to follow when writing or repairing a locator in a web test script.
When a test breaks, the repairing process generally includes three main steps: 
\begin{inparaenum}
    \item extract the cause of the breakage
    \item if a locator caused the breakage, the element is first relocated then
    \item a new locator is generated/written.
\end{inparaenum}
Beyond automated tests, this problem can also arise in more general web automation scripts covering web content crawling and \emph{Robotic Process Automation} (RPA), which heavily rely on locators to automate the navigation across web applications.

In this section, we formalize the description of two locator-related problems highlighted in Figure~\ref{fig:locator_repair}, namely the \emph{robust locator} (in blue) and \emph{locator repair} (in green) problems for the general case of web automation scripts.

\subsection{Problem Notations}
We consider that a given web page can change for various reasons, such as
\begin{inparaenum}
    \item content variation,
    \item page rendered for different regions/languages, or
    \item release of the web application.
\end{inparaenum}
No matter the cause, we distinguish $D$ and $D'$ as two versions of the same web page observed before and after a change, respectively.
More specifically, we define the following similarity notations:
\begin{compactenum}
\label{def:similarity}
    \item $D \approx D'$ if scripts written for $D$ are expected to apply on $D'$;
    \item Given 2 web elements $e \in D$ and $e' \in D'$, $e \approx e'$ if $e$ and $e'$ refer to semantically equivalent elements (\emph{e.g.}, the same menu item observed in pages $D$ and $D'$);
    \item By extension of (2), given a set of elements $E = {e_1...e_n} \subset D$ and $E' = {e'_1...e'_{n'}} \subset D'$, $E \approx E'$ if $n = n'$ and, for each $i \in [1..n]$, $e_i \approx e'_i$.
\end{compactenum}

Based on the above similarity notation, we provide the following definitions:
\begin{defn}\label{loc_def}
    Given a page $D$, and a set of elements $E = {e_1...e_n}$, the pair $(loc_{E,D}, eval)$ is a \textbf{locator} of $E$ with regard to $D$ if:
    \begin{equation}
       eval(loc_{E,D}, D) = E
    \end{equation}
    where $loc_{E,D}$ is a descriptor of $E$ and $eval$ an evaluation function that returns a set of web elements from a descriptor and an evaluation context (\emph{e.g.}, a web page).
\end{defn}
In the case of XPath-based locators, the descriptor $loc_{E,D}$ refers to an XPath query describing the elements $E$ in the page $D$ and $eval$ the XPath solver.

\begin{defn}
    Let $mut$ be a mutation function that transforms the page $D$ into another page $D'$, such as $mut(D) = D'$.
    $mut$ is said to be a \textbf{mutation} of $D$ if $D \approx D'$.
\end{defn}

\begin{defn}
    Given a locator $L = (loc_{E,D}, eval)$, $L$ is \textbf{robust} to a mutation function $mut$ if:
    \begin{equation}
       eval(loc_{E, D}, mut(D)) \approx E
    \end{equation}
\end{defn}

\subsection{Problem Statement}
Given the above definitions, we can formalize the locator problem statement along with the two following research questions.
\begin{rqn}\label{robust_locator_problem} %
    \textbf{Robust Locator}. 
    For any subset of elements on a given page $D$, how to automatically generate locators that are robust to mutations of $D$?
\end{rqn}
When evaluating a locator on a new page $D'$, the only available information to describe the targeted element is the descriptor $loc_{E,D}$, which often remains insufficient (cf. state-of-the-art techniques).

On the other hand, in the context of \textit{locator repair}, the original page $D$ from which $loc_{E,D}$ was built is available.
Thus, using definition~\ref{loc_def}, this piece of information allows to locate the originally selected elements $eval(loc_{E,D}, D) = E$.

\begin{rqn}\label{locator_repair_problem}
    \textbf{Locator Repair}. 
    Given two pages $D$, $D'$, such that $D \approx D'$ and a set of elements $E\in D$, how to locate the elements $E' \approx E$ in $D'$?
\end{rqn}
To the best of our knowledge, existing solutions to both \emph{robust locator} and \emph{locator repair} focus on the restricted case of $|E| = 1$.

Once the \textit{locator repair problem} is solved (\emph{i.e.}, $E'$ are correctly located), we need to generate new locators, which brings us back to the situation of the robust locator problem (cf. RQ.~\ref{robust_locator_problem}).

In this article, we thus present a novel approach to solve the locator repair problem.

\section{Repairing Locators with \erratum}\label{sec:implementation}
The previous section formalized both \emph{robust locator} and \emph{locator repair} problems.
The approach we report in this article, \erratum, therefore matches the DOM trees of 2 versions of a web page to solve the \emph{locator repair} problem.
Several tree matching solutions exist in the literature, such as \emph{Tree Edit Distance} (TED)~\cite{tai1979tree} or tree alignment~\cite{jiang1994alignment}.
This section therefore motivates and explains how \erratum leverages tree matching to repair locators, before discussing the choice of a tree matching implementation fitting \erratum's requirements.

\subsection{Applying Tree~Matching to Locator~Repair}
Embedding tree matching allows \erratum to leverage the tree structure in the same way an XPath-based solution would, while offering the flexibility of a more statistic-based solution.
Intuitively, a tree matching algorithm should consider all easily identifiable elements on a page (elements with rare tags, unique classes, ids, or other attributes) as \textit{anchors} to relocate less easily identifiable elements.

\begin{figure}[]
    \centering
    \includegraphics[width=.8\linewidth]{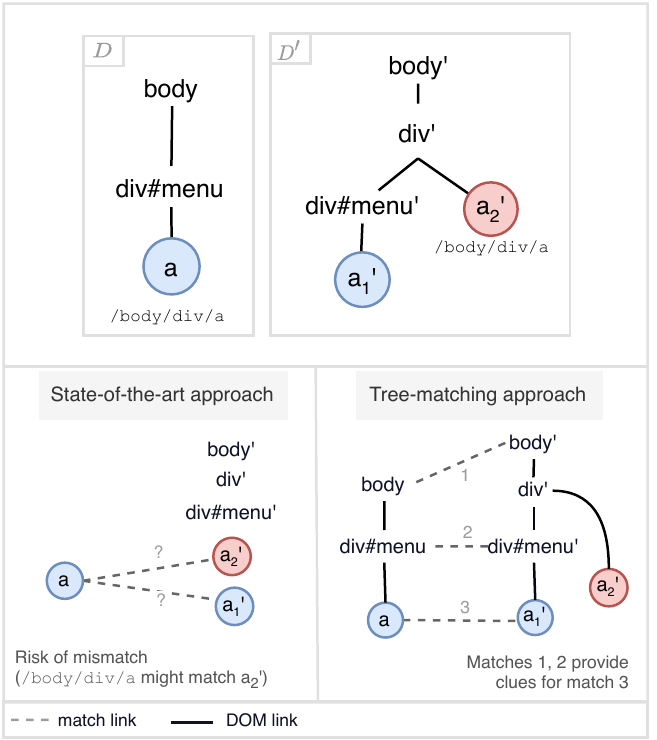}
    \caption{State-of-the-art Vs. tree matching locator repair.}
    \label{fig:holistic}
\end{figure}

Figure~\ref{fig:holistic} illustrates the benefits of a more holistic approach using tree matching.
In the example, the locator of element \texttt{a} (in blue) breaks because the mutations between $D$ and $D'$ entails a change in its absolute XPath (\texttt{/body/div/a}).
Attempting to repair such a broken locator by relying on the properties of the original element alone (state-of-the-art approaches like~\cite{choudhary2011water,stocco2018visual}) is often challenging and can easily lead to a mismatch. 
By using tree matching (cf. right-bottom of Figure~\ref{fig:holistic}), matching the parent of the element to locate (\textsf{div\#menu}) brings instead a strong contextual clue to accurately relocate the element \texttt{a$_1$'} whose locator was broken~\cite{brisset2020sftm}.

Formally, given a pair of page versions $D$ and $D'$, we:
\begin{compactenum}
    \item parse $D$ and $D'$ into DOM trees $T$ and $T'$.
    Consequently, $nodes(T)$ is the set of elements in the DOM tree $T$;
    \item apply tree matching to $T, T'$ yielding a matching $M \subset nodes(T) \times nodes(T')$.
    If the resulting matching $M$ is accurate, then $\forall (e, e') \in M, e \approx e'$;
    \item use the resulting matching $M$ to repair the broken locator(s).
\end{compactenum}

Regarding the test repair process illustrated in Figure~\ref{fig:locator_repair}, our approach thus fits in the block "\textsf{relocate element}" (in green) by matching the elements of $D$ in $D'$ and reporting the relocated element. 
Thus, once the element is relocated using tree matching---\emph{i.e.} \erratum found $e' \in D'|e \approx e'$---we only need to generate a new locator $loc_{e', D'}$ to achieve the test repair process.
This task can be performed using solutions to the robust locator problem, like ROBULA~\cite{leotta2014reducing}, and is therefore considered as out of the scope of this article.

\subsection{Integrating a Scalable Tree Matching Algorithm}
The state-of-the-art approach to match two trees is \emph{Tree Edit Distance} (TED)~\cite{tai1979tree}. 
When comparing two trees $T$ and $T'$, TED-based approaches rely on finding the optimal sequence of relabels, insertions and deletions that transforms $T$ into $T'$.
Unfortunately, TED might be unsuitable to match real-life web pages due to two core restrictions~\cite{Kumar2011_FTM}:
\begin{inparaenum}
    \item if two nodes $e$ and $e'$ are matched, the descendants of $e$ can only match with the descendants of $e'$, and
    \item the order of siblings must be preserved.
\end{inparaenum}
Furthermore, TED is computationally expensive ($O(n^3)$ for the worst-case complexity~\cite{bringmann2018tree}) and, more practically, our preliminary experimentation has shown that applying the state-of-the-art implementation of TED, named APTED~\cite{pawlik2016tree}, on the \textit{YouTube} page takes several minutes. 
We believe that, in addition to qualitative restrictions, such computation times are not acceptable when periodically repairing locators on real websites.

Further studies of TED proposed to improve computation times~\cite{jiang1995alignment, valiente2001efficient, zhang1996constrained}, but at the cost of even more restrictive constraints on the produced matching (\emph{e.g.}, the tree alignment problem~\cite{jiang1995alignment} restricts the problem to transformations where insertions are performed before deletions).

To the best of our knowledge, the only contribution that provides a solution to the general (restriction-free) tree-matching problem is the \emph{Flexible Tree Matching} (FTM) algorithm~\cite{Kumar2011_FTM}.
FTM models tree matching as an optimization problem: given two trees $T$ and $T'$ how to build a set of pairs $(e,e') \in T \times T'$ such that the similarity between all selected node pairs is maximal.
The similarity used by FTM combines both the labels and the topology of the tree.

However, as shown in~\cite{brisset2020sftm}, the theoretical complexity of FTM is high ($O(n^4)$) and the implementation of FTM was shown to take more than an hour to match a web page made of only 58 nodes, while the average number of nodes on a web page observed in our dataset is $1,507$.
Consequently, we believe that such computation times make FTM unpractical in the context of locator repair.

\subsection{Matching DOM Trees by Similarity}\label{sec:SFTM}
Given the limitation of FTM, \erratum{} implements a \emph{Similarity-based Flexible Tree Matching} (SFTM) algorithm, which is an extension of state-of-the-art FTM to improve the computation times of FTM without any restriction on the resulting matching. 

In the context of \erratum{}, our SFTM algorithm assumes that, given a web page, several elements are easily identifiable by considering their intrinsic properties.
Our algorithm, therefore, first assigns scores to all possible matches between nodes from the two trees based on their label and only then uses the topology of the trees to adjust these scores.
We walk through the key steps of the SFTM algorithm we implemented in \erratum{} by using HTML snippets reported in Figure~\ref{fig:example_html}.
The figure provides two versions ($D$ and $D'$) of a simplified HTML code sample extracted from the homepage of the famous search engine \textit{duckduckgo.com}.
In this example, our purpose is to relocate $\text{a}_1 \in D$ with $\text{a}_1' \in D'$.

\begin{figure}
    \centering
    \begin{subfigure}[b]{\linewidth}
        \centering
        \caption{Original document $D$.}
        \begin{lstlisting}[language=html, label={fig:first_version}]
<div class="content-info__item"> /*!\mymk{div_1}!*/
    <div class="item__title">...</div> /*!\mymk{div_2}!*/
    <div class="item__subtitle"> /*!\mymk{div_3}!*/
        ... 
        <a href="/plugins">Plugins</a> /*!\mymk{~a_1~}!*/
    </div>
</div>
        \end{lstlisting}
    \end{subfigure}
    \hfill
    \begin{subfigure}[b]{\linewidth}
        \centering
        \caption{Updated document $D'$ from $D$.}
        \begin{lstlisting}[language=html, label={fig:second_version}]
<div class="items-wrap"> /*!\mymk{div'_4}!*/
    <div class="item"> /*!\mymk{div'_1}!*/
    <div class="item__title">...</div> /*!\mymk{div'_2}!*/
        <div class="item__subtitle"> /*!\mymk{div'_3}!*/
            ... 
            <a href="/extensions">Extensions</a> /*!\mymk{~a'_1~}!*/
        </div>
    </div>
    <div> /*!\mymk{div'_5}!*/  
        ...
        <a href="/newsletter">Newsletter</a> /*!\mymk{~a'_2~}!*/
    </div>
</div>
        \end{lstlisting}
    \end{subfigure}
    \caption{Two versions of an HTML snippet extracted from the homepage of \emph{duckduckgo.com}.}
    \label{fig:example_html}
\end{figure}

\begin{figure*}[!t]
    \centering
    \includegraphics[width=\linewidth]{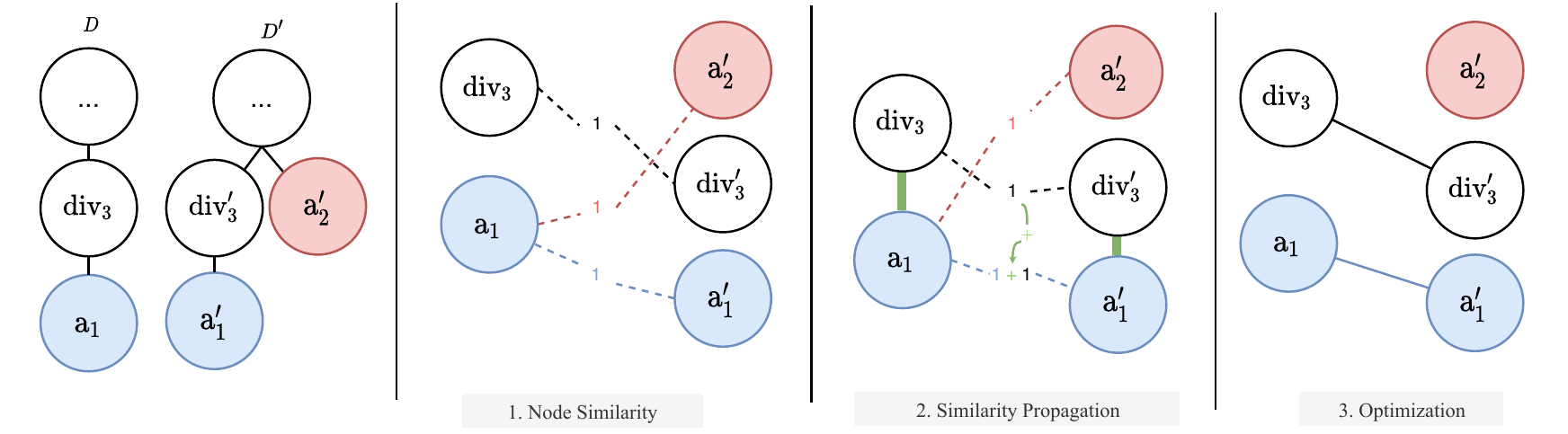}
    \caption{Key steps followed by our \emph{Similarity-based Flexible Tree Matching} (SFTM) algorithm.}
    \label{fig:steps_sftm}
\end{figure*}

Unlike the state-of-the-art matching algorithms, SFTM first tries to match elements in $D'$ whose labels are similar to $D$, before using these matched elements to adjust the similarity of surrounding elements in the tree.
For example, the \emph{similarity scores} of the tuple $(a_1,a'_1)$ links will increase as their direct parents $(div_3,div'_3)$ are matched with confidence.
Figure~\ref{fig:steps_sftm} summarizes the SFTM algorithm's key steps and the remainder of this section provides an overview of its integration in \erratum{} (cf. green box in Figure~\ref{fig:locator_repair}).
The interested reader can refer to our technical report~\cite{brisset2020sftm} for an exhaustive description of our SFTM algorithm.

\paragraph{Step\,1: Node Similarity}
Elements of DOMs $D$ and $D'$ are compared. 
The first step consists in computing an \emph{initial similarity} $s_0 : D \times D' \to [0, 1]$.
For each pair of nodes $(e, e') \in D \times D'$, $s_0(e,e')$ measures how similar the labels of $e$ and $e'$ are.
In HTML pages, the label of a node $e \in D$ that we use is the set of tokens obtained from applying a tokenizer to the HTML code describing $e$. 
This may include the type of the HTML element, its attributes, and eventually the raw content associated to this element---\emph{i.e.}, thus ignoring the content from the child elements.
\begin{ex}
The \emph{label} computed for $div_1$ (cf. Figure~\ref{fig:example_html}) includes the following tokens: $\{\texttt{div}, \texttt{class}, \texttt{content-info\_\_item} \}$. %
\end{ex}

To compute $s_0$, SFTM first indexes the labels of each node of $D$.
The idea of this step is to prune the space of possible matches by pre-matching nodes with similar labels.
When indexing, to improve the accuracy of $s_0$, we apply the \emph{Term Frequency-Inverse Document Frequency} (TF-IDF)~\cite{jones1972statistical} formula to take into account how rare each token is.
\begin{ex}
    Following our previous Example~\ref{fig:example_html}, when considering the match $(div_1, div'_1)$:
    \begin{compactenum}
        \item token $\texttt{div}$ will yield very few similarity points, since it is included in the labels of almost all the nodes,
        \item token $\texttt{content-info\_\_item}$ will increase the score significantly, as it only appears once in both documents.
    \end{compactenum}
\end{ex}

In general, very common tokens bring very few information to the relevance of a given match, while they cause a significant increase of potential matches to consider.
That is why, in order to reduce the computation times, the algorithm rules out most common tokens.

\paragraph{Step\,2: Similarity Propagation}
The initial similarity $s_0$ only takes into account the labels of nodes.
In this second step, the idea is to enrich the information contained in $s_0$ by leveraging the topology of the trees $D$ and $D'$.
\begin{ex}
    In Example~\ref{fig:example_html}, it is hard to choose the correct match $m_1 = (a_1, a'_1)$ over the incorrect one $m_2 = (a_1, a'_2)$ by only considering labels, since all three elements share the same set of tokens: $\{\texttt{a}, \texttt{href}\}$.
    In the similarity propagation step, we leverage the fact that the parents of $a_1$ and $a'_1$ are similar to increase the similarity between $a_1$ and $a'_1$, thus preferring $m_1$ over $m_2$.
\end{ex}
In general, for each considered match $(e, e') \in D \times D'$, the parents of $e$ and $e'$ gets more similarity points if $e$ and $e'$ are similar and inversely, $s(e, e')$ is increased if the parents of $e$ and $e'$ are similar (with regard to $s_0$).
We call $s$ the final similarity produced by this step.

\paragraph{Step\,3: Optimization}
Producing the optimal matching with regards to the computed similarity means selecting the full set of matches such that each element of $D$ is matched with at most one element of $D'$ and the sum of similarity scores of the selected matches is maximum.

In order to approximate the optimal set of matches, SFTM implements the Metropolis algorithm~\cite{metropolis1953equation}.
The idea is to randomly walk through several possible configurations (set of matches) to converge towards the optimal one.

At the end of the optimization step, the SFTM algorithm yields a full matching $M \subset D \times D' $ comprising matches between nodes of $D$ and $D'$.
These matches can be analyzed by \erratum{} to locate broken locators and fix them by generating new locators in the target document $D'$.

\section{The Robust Locator Benchmark}\label{sec:benchmark}
In the context of this article, we are interested in covering the following research questions:
\begin{rqn}\label{rq:performance}
    How does \erratum{} perform in solving the locator repair problem (cf. RQ.~\ref{locator_repair_problem}) when compared to state-of-the-art solutions?
\end{rqn}
\begin{rqn}\label{rq:influenceFactors}
    What are the factors influencing the accuracy of WATER and \erratum?
\end{rqn}
\begin{rqn}\label{rq:computationTime}
    How quickly can \erratum{} repair broken locators when compared to state-of-the-art solutions?
\end{rqn}
This section, therefore, describes the benchmark we propose to assess these questions.

\subsection{Evaluated Locator Repair Solutions}\label{sec:consideredSolutions}
We compare two solutions: 
\begin{inparaenum}
    \item \erratum{}, our approach to repair broken locators by leveraging flexible tree matching, and
    \item WATER, the reference implementation of a locator repair technique applied to web test scripts~\cite{choudhary2011water}.
\end{inparaenum}

The original algorithm of WATER analyses a given test case, finds the origin of the test breakage, and suggests potential repairs to the developer.
In the context of this article, we are interested in the most challenging part of the algorithm: the part that repairs broken locators, if needed.
Given the originally located element $e \in D$, WATER attempts to find $e' \in D'$ such that $e' \approx e$ by scanning over all elements in $D'$ such that $tag(e') = tag(e)$ and selecting the elements most similar to $e$. 
The similarity between two elements $e_1, e_2$ used by WATER mostly consists in computing the Levenshtein distance between the absolute XPaths of both elements ($Levenshtein(XPath(e_1), XPath(e_2))$) combined with other element properties similarity (\emph{e.g.}, visibility, z-index, coordinates).
In our evaluation, we re-implemented this part of the WATER algorithm to compare its performance to \erratum{}.

We initially considered VISTA~\cite{stocco2018visual} as a baseline, even though the approach they use (computer vision) is radically different from \erratum{} and WATER.
However, despite our efforts, we failed to run their implementation and received no answer when trying to contact the authors. 

Note that, in this evaluation, we focus on \textit{single-element locator} cases of the locator repair problem (we only try to repair \textit{single-element locators}), which is the worst-case scenario for \erratum{}.
The reasons for this decision are:
\begin{inparaenum}
    \item The state-of-the-art solutions to both repair and robust locator problems only treat this case and in particular, WATER can only repair locators locating a single element,
    \item \erratum{} reasons on the whole trees, so locating several independent elements is done the same way as locating a group of elements.
\end{inparaenum}

\subsection{Versioned Web Pages Datasets}
In the remainder of this article, we propose two datasets to compare \erratum{} and WATER against potential evolutions of web pages.
Given two versions of the same page $D, D'$, and a set of elements $E \subset D$, the locator repair problem consists in locating a set of elements $E' \subset D'$, such that $E' \approx E$.
To evaluate the performance of a locator repair tool, we thus need what we call a \textbf{DOM versions dataset}: a dataset of pairs $(D, D')$, such that $D \approx D'$.

A DOM version dataset is also required to evaluate solutions to the robust locator problem.
To build such a dataset, previous works on locator repair~\cite{leotta2016robula+,leotta2014reducing} and robust locator~\cite{stocco2018visual,choudhary2011water,hammoudi2016waterfall} manually analyzed different versions of a few open source applications (like Claroline, AddressBook or Joomla).
These evaluations are significantly limited in size (never beyond a dozen of websites considered) and hard to reproduce since the exact versions of the open source applications used are often not provided or available.

In our study, we therefore introduce the first large-scale, reproducible, real-life \textit{DOM versions dataset} that can be used to assess locator repair solutions, and is composed of two parts:
\begin{compactenum}
    \item A {\bf \textsc{Mutation} dataset}~\cite{brisset2020sftm} generated by applying random mutations to a given set of web pages (see Section~\ref{mutationDataset}),
    \item A {\bf \textsc{Wayback} dataset} collects past versions of popular websites from the Wayback API (see Section~\ref{waybackDataset}).
\end{compactenum}

Then, for each pair $(D, D')$ in the dataset, our experiments consist of selecting a set of elements to locate in $D$ and in comparing both \erratum{} and WATER trying to find the corresponding element on $D'$.

Table~\ref{table:describing_datasets} describes both datasets in terms of:
\begin{compactenum}
    \item \textbf{\# Unique URLs}: the number of unique URLs among the total of version pairs in the dataset. The duplication is due to the fact that there can be several mutations or successive versions of the same web page. In the case of the {\sc Wayback} dataset, more popular websites are more represented (see Section~\ref{waybackDataset});
    \item \textbf{\# Version pairs}: the number of considered pairs of web pages $(D, D')$,
    \item \textbf{\# Located elements}: the number of elements $e \in D$ that any solution should locate in $D'$.
\end{compactenum}

\begin{table}[!htbp]
    \caption{Description of the \textsc{Mutation} \& \textsc{Wayback} datasets.}
    \label{table:describing_datasets}
    \centering
    \resizebox{.6\linewidth}{!}{%
        \begin{tabular}{|l|r|r|}
        \hline
         Dataset      & {\sc Mutation} & {\sc Wayback} \\ \hline
        \hline
        \# Unique URLs      &      650 &      64 \\
        \# Version pairs    &    3,291 &   2,314 \\
        \# Located elements &   49,305 &  34,421 \\
        \hline
        \end{tabular}
    }
\end{table}

The two datasets we provide are complementary. 
Since the {\sc Mutation} dataset is generated by mutating elements from an original DOM $D$, the ground truth matching between $D$ and its associated mutation $D'$ is known to easily evaluate the solution on a very large amount of version pairs. 
However, since the versions are artificially generated, this dataset is synthetic and, as such, might not entirely reflect the actual distribution of mutations happening along a real-life website lifecycle.

Then, the {\sc Wayback} dataset is composed of real website versions mined from the Wayback API: an open archive that crawls the web and saves snapshots of as many websites as possible at a rate depending on the popularity of the website.\footnote{\url{https://archive.org/help/wayback_api.php}}
In the {\sc Wayback} dataset, mutations between $D$ and $D'$ are not synthetic, but as a result, the ground truth matching between $D$ and $D'$ is unknown.
In our evaluation, we thus had to manually label a sample of the results obtained on this dataset, which limits the scalability of the experiment compared to the {\sc Mutation} dataset. 

The following sections provide more details on how both datasets were built.

\subsubsection{Building the {\sc Mutation} dataset.}\label{mutationDataset}
We extend the technique we introduced to generate a \textsc{Mutation} dataset in~\cite{brisset2020sftm}.
The mutation dataset is built by applying a random amount of random mutations to a set of original webpages: for each original DOM $D$, 10 mutants are created by applying mutations to $D$.
Since the mutations applied to $D$ to construct each mutant $D'$ are known, the ground truth matching between $D$ and $D'$ is also known.
Knowing the ground truth matching on the mutation dataset allows us to evaluate our locator repair solution on a very large dataset. 
Table~\ref{table:mutations_used} describes the set of DOM mutations that can be observed along evolution of a web page.

\begin{table}
    \caption{Mutations applied in the \textsc{Mutation} dataset~\cite{brisset2020sftm}.}
    \label{table:mutations_used}
    \centering
    \resizebox{.8\linewidth}{!}{%
        \begin{tabular}{|c|l|}
           \hline
           \bf Type       & \bf Mutation operators\\
           \hline
           \hline
           \em Structure & \tt remove, duplicate, wrap, unwrap, swap \\ \hline
           \em Attribute & \tt remove, remove words \\ \hline
           \em Content   & \tt replace with random text, change letters, \\
                         & \tt remove, remove words \\
           \hline
        \end{tabular}
    }
\end{table}

The original websites from which mutants were generated were randomly selected from the Top\,1K Alexa.
Figure~\ref{fig:distribution} depicts the distribution of DOM sizes in this synthetic dataset. 

\begin{figure}[h]
  \centering
  \includegraphics[width=.8\linewidth]{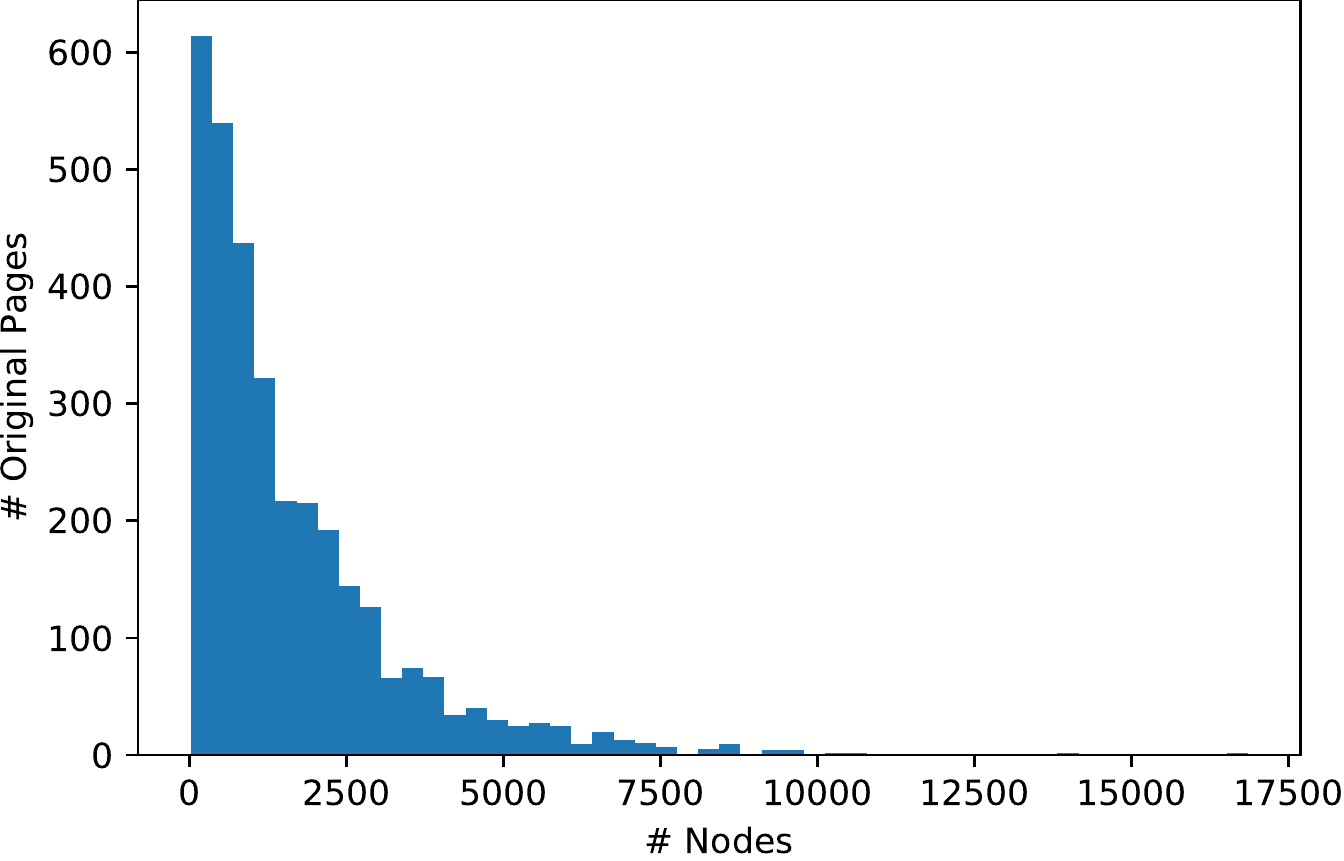}
  \caption{Distribution of DOM sizes (in number of nodes) in the \textsc{Mutation} dataset.}
  \label{fig:distribution}
\end{figure}

This dataset was built with an automation tool that we made available along with its source code~\ref{sec:conclusion}.
From a given list or source URLs, our tool creates a dataset of randomly mutated webpages following the above-described approach.

\subsubsection{Buidling the {\sc Wayback} dataset.}\label{waybackDataset}
This dataset encloses a list of $(D, D')$ DOM pairs where $D$ and $D'$ are two versions of the same page (\emph{e.g.}, \textit{google.com} between 01/01/2013 and 01/02/2013).
Two versions can be separated by different gaps in time.
In this section, we explain how we used the Wayback API to build this dataset.
The Wayback API can be used to explore past versions of websites.
The two endpoints we used to build the dataset can be modeled as the following functions:
\begin{align*}
versionsExplorer &::& (url, duration)  & \to & timestamp[] \\
versionResolver  &::& (url, timestamp) & \to & document
\end{align*}
The \textit{versionsExplorer} retrieves the list of available snapshots between two dates, while the \textit{versionResolver} returns the snapshot of a given \textit{url} at the requested timestamp.

Using these endpoints, for each website URL considered, we:
\begin{compactenum}
\item retrieved the timestamps of all versions between 2010 and today using the \textit{VersionExplorer},
\item generated a list of all pairs of timestamps with one of following differences in days ($\pm 10\%$): $\text{[7, 15, 30, 60, 120, 240, 360]}$,
\item picked up to $1,000$ random elements from the list of timestamps pairs,
\item resolved each selected timestamp pair using the \textit{versionResolver}.
\end{compactenum}

Similarly to the {\sc Mutation} dataset, the URLs we fed to our algorithm were taken from the Top\,1K Alexa.
Since both datasets are based on the same set of URLs (taken from Alexa), the distribution of the {\sc Wayback} dataset is very similar to the {\sc Mutation} one (cf. Figure~\ref{fig:distribution_wayback}).

\begin{figure}[h]
  \centering
  \includegraphics[width=.8\linewidth]{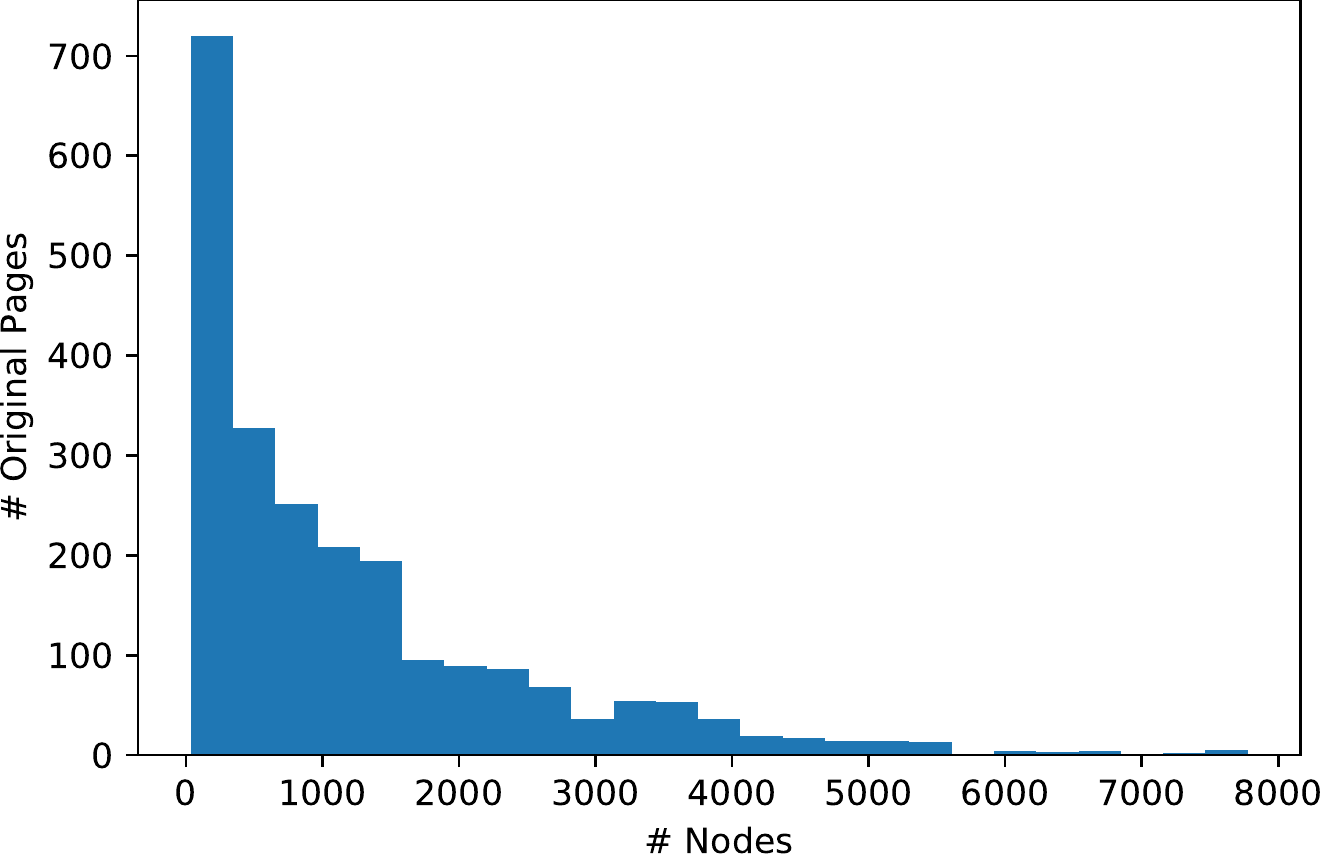}
  \caption{Distribution of DOM sizes (in number of nodes) in the \textsc{Wayback} dataset.}
  \label{fig:distribution}
\end{figure}

\subsubsection{Selecting the elements to repair.}
\erratum{} and WATER operate in different ways.
\erratum{} takes two trees $(D, D')$ and returns a matching between each element of the trees, thus solving any possible broken locator between $D$ and $D'$.
The algorithm extracted from WATER is a more straightforward solution to the locator repair problem as formally described (cf. Section~\ref{locator_repair_problem}): it takes a pair of DOM versions $(D,D')$ and an element $e \in D$ as input and returns an element $e' \in D'$ (or $null$ if it fails to find any candidate for the matching).

Consequently, in the case of the WATER algorithm, the following question arises: given a pair $(D, D')$ taken from the DOM version dataset, which elements of $D$ should be picked for repair? 
Ideally, we would try to locate every element of $D$ in $D'$ to obtain a comprehensive comparison with \erratum.
Unfortunately, the computation times of WATER make it impractical to locate every single element from $D$ in $D'$.
Selecting realistic targets for locators is a non-obvious task since many elements in the DOM would not be targeted in a test script (\emph{e.g.}, large container blocks, invisible elements, aesthetic elements).
Therefore, for each version pair, we randomly select up to $15$ clickable elements from $D$.
We focus on clickable elements as this is the most common use case for web UI testing (to trigger interactions), and WATER has specific heuristics to enhance its accuracy on links.
By considering clickable elements, we
\begin{inparaenum}
    \item make sure to choose realistic elements, and
    \item compare to WATER on its most typical use case.
\end{inparaenum}

Regarding the sample size, considering $15$ elements per web page leads to selecting 34,000+ elements in both datasets.
As the average number of nodes per web page in each dataset is around $1,500$, this means that there are more than 3.6M candidate locators for repair in each dataset.
Therefore, the confidence interval at 95\,\% of the measurements applied to the 34K sample of located elements is 0.5\,\%. 

\subsection{Evaluating of the Matched Elements}\label{sec:manualLabeling}
On the {\sc Mutation} dataset, the signatures attributes are preserved after mutations (but ignored when applying either locator repair solution), thus providing the ground truth matching between the DOMs of a version pair.

For the {\sc Wayback} dataset though, this information is not available.
For each version pair $(D, D')$, the evaluation of both solutions yields to a list of suggested matching $(e, e'_{ERRATUM})$ and $(e, e'_{WATER})$ where $e \in D$ and $e'_{ERRATUM}, e'_{WATER} \in D'$.
In both cases, $e'$ may be null in case no matching was found.
Given the above situation, the labeling process consists of determining whether the matching element of $e$ is $e'_{ERRATUM}$, $e'_{WATER}$, or neither.
In many cases, $e'_{ERRATUM} = e'_{WATER}$ (consensus).
We choose to focus our manual labeling effort on cases where WATER and \erratum{} disagree and assume that both solutions are right otherwise.

\begin{figure*}
  \begin{center}
  \includegraphics[width=1.1\linewidth]{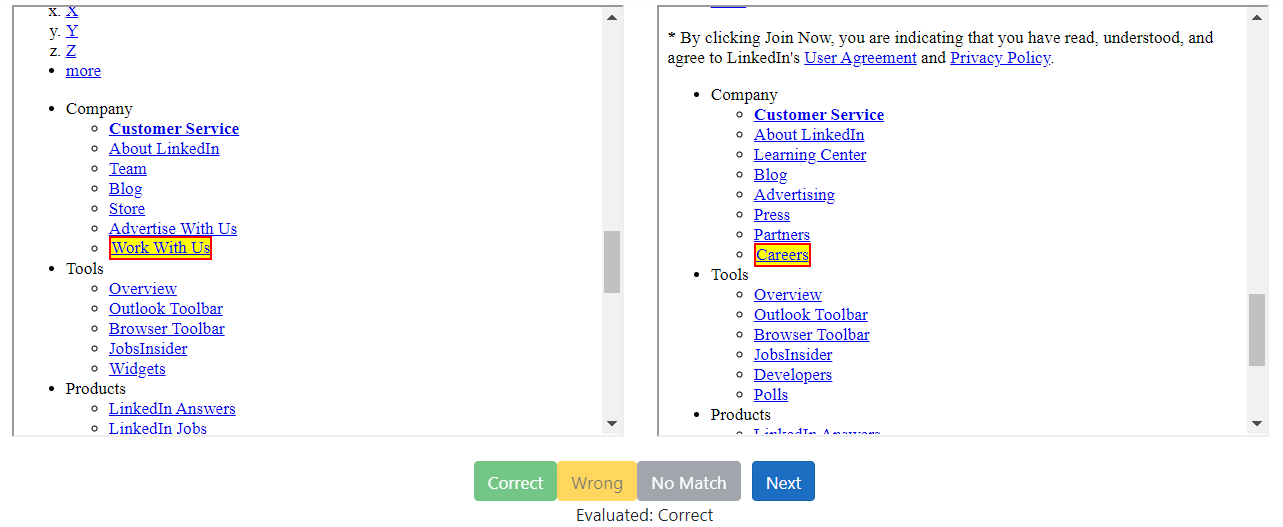}
  \caption{Labeling a given element matched by \erratum{} on two versions of the \textsf{Linkedin} homepage. The screenshot comes from the visual matching application we created to manually label disagreements between \erratum{} \& WATER.}
  \label{fig:disagreement}
  \end{center}
\end{figure*}

Thus, to label the disagreements between \erratum{} and WATER, we developed a web application (cf. Figure~\ref{fig:disagreement}) to display the identified elements on both versions of the DOM version pair and label the matching as either correct or wrong.

When we defined the similarity equivalence between two elements (cf. Definition~\ref{def:similarity}), we mentioned the potential subjective part of the measure.
To lessen this subjective part and label the proposed matchings as objectively as possible, we systematically recommended the following guidelines:
\begin{compactenum}
    \item Sometimes, matched elements are not visible (it happens when the visibility of some parts of the page is triggered dynamically). In this case, if elements in both versions are not visible, the locator is skipped, otherwise, the matching is considered as \textsf{mismatch};
    \item Sometimes, a link appears in different locations on the website (often sign-in links). Matching two such links from different locations is considered wrong even though the two links might be assumed to have a similar semantic value. Therefore, we always consider the surrounding of the located element to judge whether the matching is \textsf{correct} or \textsf{mismatch};
\end{compactenum}

\section{Empirical Evaluation}\label{sec:evaluation}
This section evaluates locator repair solutions along with two criteria, accuracy and performance, to answer our research questions.

\subsection{Evaluation of Repair Accuracy}
In this section we answer RQ.\ref{rq:performance}: 
How does \erratum perform in solving the locator repair problem (RQ.\,\ref{locator_repair_problem}) when compared to state-of-the-art solutions?

\begin{figure}[]
  \centering
  \includegraphics[width=.7\linewidth]{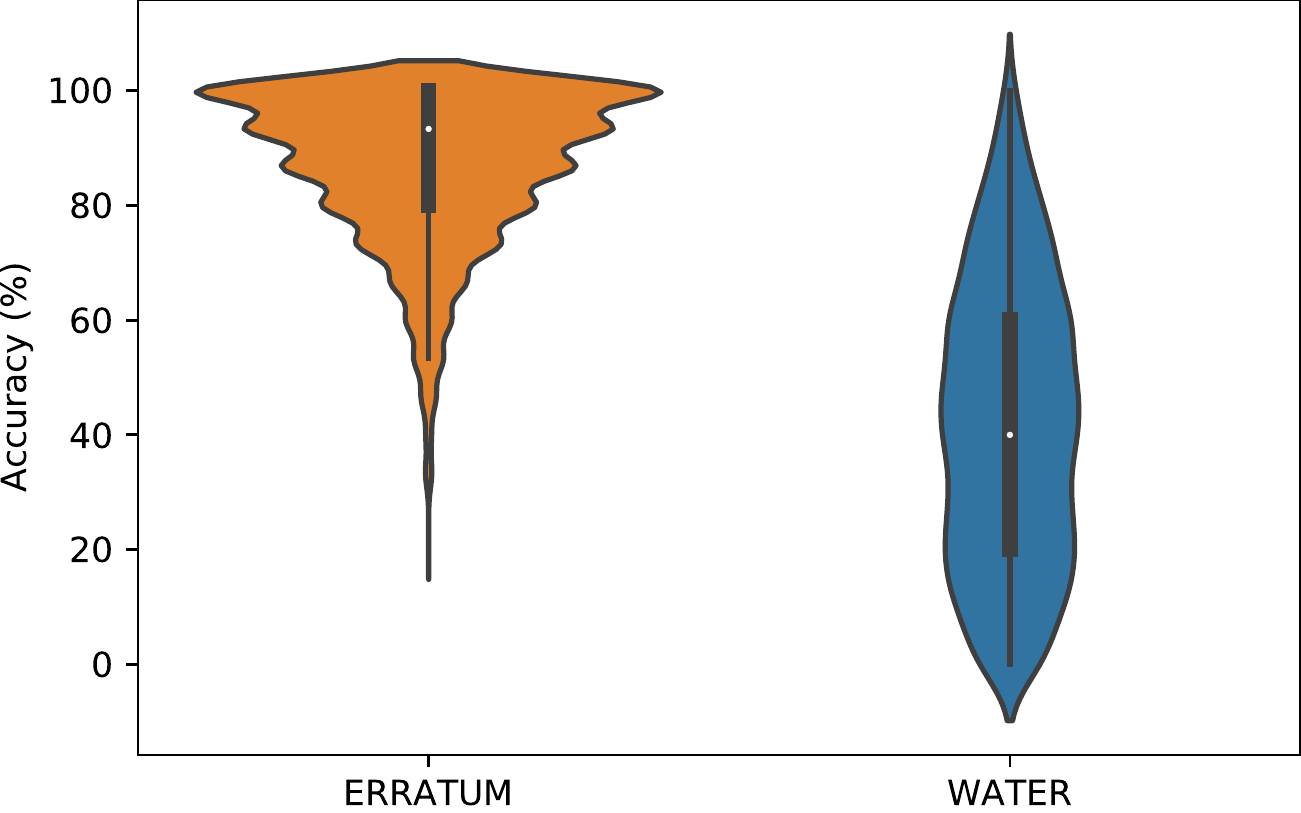}
  \caption{Accuracy distribution of \erratum{} and WATER on the \textsc{Mutation} dataset.}
  \label{fig:mutation_boxplot}
\end{figure}

\vspace{6pt}\noindent{\bf Repair accuracy on the \textsc{Mutation} dataset.}
Figure~\ref{fig:mutation_boxplot} summarizes the distribution of the accuracy of \erratum{} and WATER as a violin plot over the $3,291$ version pairs of our {\sc Mutation} dataset.
For each version pair $(D, D')$, the reported accuracy ratio corresponds to the ratio of the $15$ selected elements from $D$ that are accurately located in $D'$. The figure shows

There are two ways a repair solution can fail to locate an element $e \in D$ in $D'$: 
\begin{inparaenum}
    \item a \textsf{mismatch}, when the original element $e \in D$ has been matched to the wrong element $e' \in D'$, or
    \item a \textsf{no-match}, when the algorithm does not manage to locate $e$ in $D'$. 
\end{inparaenum}
In case of failure, a \textsf{no-match} is always preferred to a \textsf{mismatch}, since a \textsf{no-match} alerts the developer about failure.
Thus, considering the two classes of errors on the {\sc Mutation} dataset, Table~\ref{tab:mismatch} summarizes the ratio of \textsf{no-match} and \textsf{mismatch} reported by both solutions.
In particular, the data shows a significant advantage in favor of \erratum when it comes to reducing locator mismatches, compared to WATER.

\begin{table}[H]
    \caption{Errors distribution of \erratum{} and WATER on the \textsc{Mutation} dataset.}\label{tab:mismatch}
    \centering
    \resizebox{.6\linewidth}{!}{%
        \begin{tabular}{|c|rr|rr|}
        \cline{2-5}
        \multicolumn{1}{l|}{ } & \multicolumn{2}{c|}{\bf \erratum} & \multicolumn{2}{c|}{\bf WATER} \\
        \cline{2-5}
        \hline
        \textsf{correct} & $42,876$ & $(87.0\%)$ & $20,740$ & $(42.1\%)$ \\
        \textsf{mismatch} & $4,420$ &  $(9.0\%)$ & {\color{red}$26,820$} & {\color{red}$(54.4\%)$} \\
        \textsf{no-match} & {\color{editorGreen}$2,009$} & {\color{editorGreen}$(4.0\%)$} & $1,745$ & $(3.5\%)$ \\
        \hline 
        \hline
        \bf Total: & $49,305$ & $(100\%)$ & $49,305$ & $(100\%)$ \\
        \hline 
        \end{tabular}%
    }
\end{table}

To further understand which factors influence the accuracy of \erratum and WATER (RQ.\ref{rq:influenceFactors}), we studied the evolution of accuracy according to three factors:
\begin{inparaenum}
    \item the type of mutations applied,
    \item the size of the DOM (number of nodes) of the original page $D$, and
    \item the mutation ratio applied to the original page $D$ to obtain the mutant $D'$.
\end{inparaenum}

\begin{figure}[]
  \centering
  \includegraphics[width=.8\linewidth]{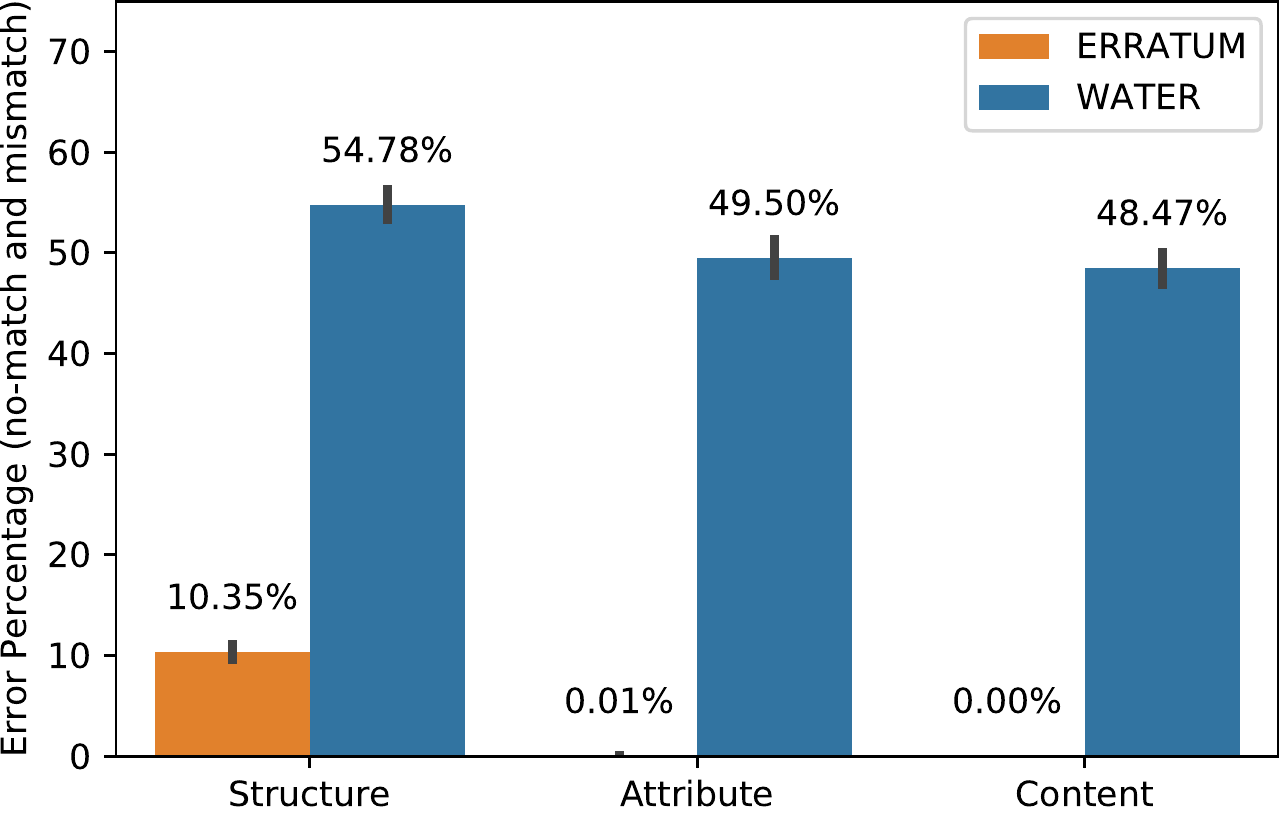}
  \caption{Error percentage according to the mutation type.}
  \label{fig:mutationTypeAnalysis}
\end{figure}

First, to assess the impact of the mutation type on the accuracy of \erratum and WATER, we used a constrained version of the \textsc{Mutation} dataset with only one mutation operation applied for each mutant. 	
In the original \textsc{Mutation} dataset, a mutant $D'$ of a page $D$ is obtained by picking a random number $l$ of random nodes $n_1, n_2...n_l \in D$ and applying a random mutation type (cf. Table~\ref{table:mutations_used}) to each node.
In the constrained version, we use the same original pages $D$, but select a single random mutation operation per mutant $D'$.
We then apply the mutation operation to $l$ randomly selected nodes: $n_1, n_2...n_l$. 
For each original page $D$, the result is a list of mutants such that each mutant $D'$ was obtained using only one mutation operation on a random amount of random nodes.
Figure~\ref{fig:mutationTypeAnalysis} depicts the sensitivity of both locator repair solutions on this alternative dataset.
The vertical lines on top of each bar represent the confidence interval.
The figure highlights that \erratum is almost exclusively sensitive to structural mutations.
In particular, the average accuracy of \erratum is not sensitive to content mutations on the page, which is expected since the algorithm ignores the content of the nodes by default.
The very low sensitivity of \erratum to attributes related mutations is more surprising as attributes account for a major part of the similarity metric of the algorithm.
For this reason, we believe that the mutation of attributes might have more impact when combined with structural mutations, which does not happen in the constrained \textsc{Mutation} dataset.

\begin{figure}[]
  \centering
  \includegraphics[width=.8\linewidth]{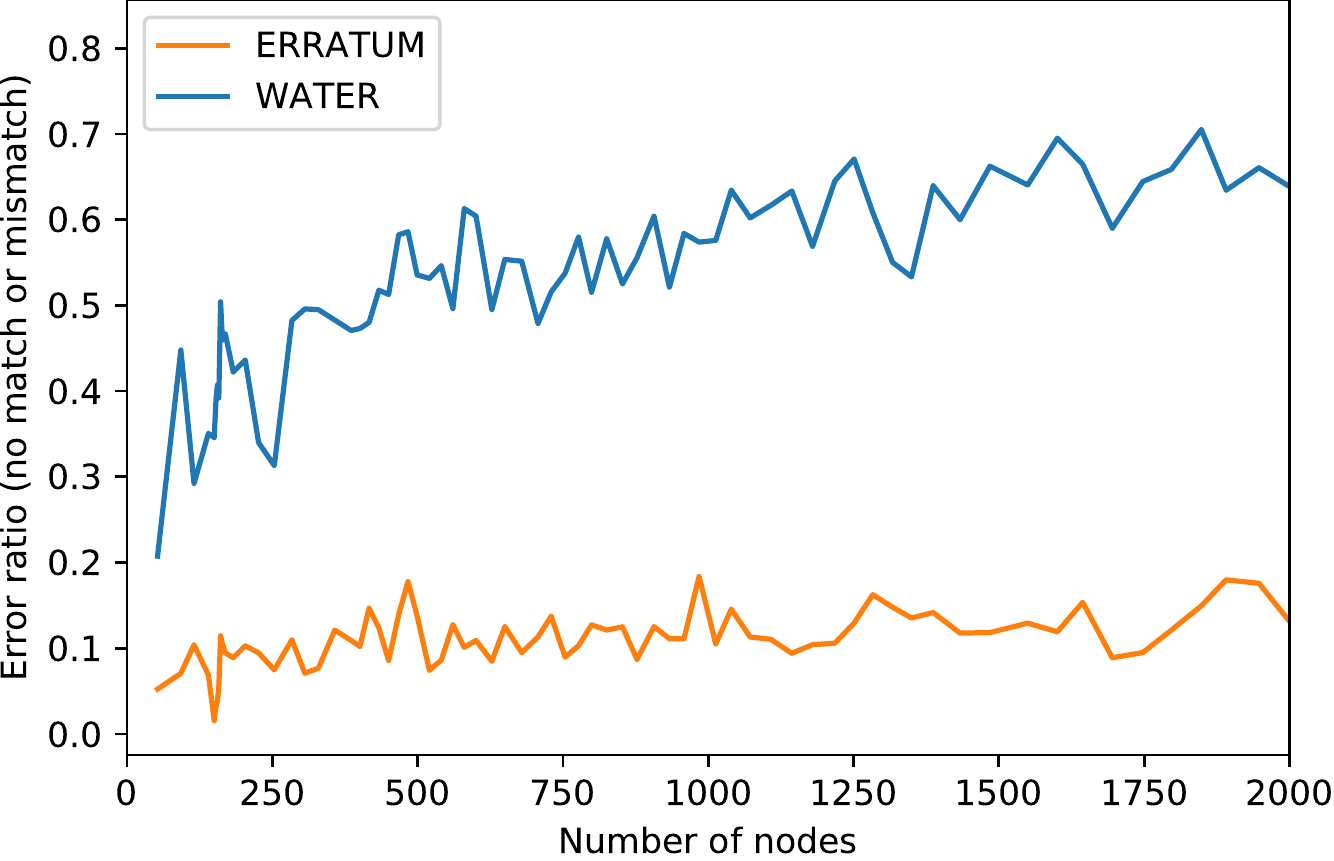}
  \caption{Errors rate evolution according to DOM size.}
  \label{fig:nbNodesAccuracy}
\end{figure}

Then, regarding the impact of the size of the DOM, our analysis concludes that WATER loses accuracy when the number of nodes increases (cf. Figure~\ref{fig:nbNodesAccuracy}), while \erratum exhibits a more stable performance.
We believe this is a key insight in understanding the limitation of WATER when compared to \erratum.
For each element $e \in D$ to locate, WATER searches through all same-tag elements in $D'$ (the \textit{candidates}) and picks the closest one to $D$, with respect to WATER's chosen similarity metric.
We believe that the sensitivity of WATER to the number of nodes comes from the fact that the number of \textit{candidate} matchings for a given element $e$ tends to grow with the size of the DOM, which increases the complexity of the ordering-by-similarity task.
Conversely, additional nodes provide more "anchor" points to \erratum, partially compensating the increase in possible combinations.

\begin{figure*}[]
  \centering
  \includegraphics[width=1\linewidth]{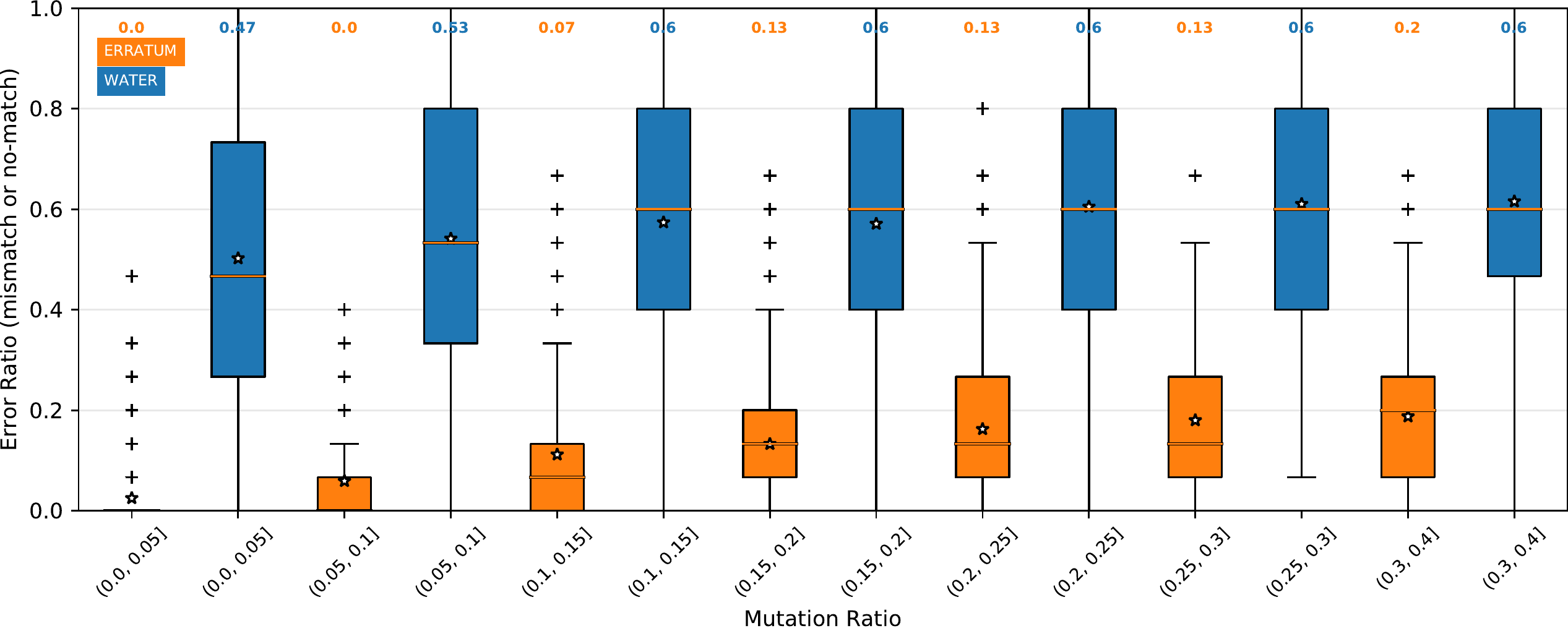}
  \caption{Errors rate evolution according to the mutation ratio.}
  \label{fig:erroRatioMutation}
\end{figure*}

Finally, regarding the impact of the mutation ratio ($\frac{\#mutations}{\#nodes}$), Figure~\ref{fig:erroRatioMutation} reports on how \erratum and WATER's errors evolve when increasing the number of mutations ($\#mutation$) on the original page $D$.
The figure contains more information than most common box plots, in particular: the stars indicate the average ratio, the horizontal orange lines, the medians whose values also appear above the boxes.
As expected, both solutions lose accuracy when the mutation ratio increases, but one can still observe that \erratum demonstrates a significant advantage over WATER, no matter the mutation ratio, and exhibiting only 20\% of errors on average (against 67\% for WATER) when the ratio of mutation exceeds 20\% of the nodes.

\vspace{6pt}\noindent{\bf Repair accuracy on the \textsc{Wayback} dataset.}
Since the {\sc Wayback} dataset does not provide any ground truth matching, we had to manually label the results of the evaluation.
We ran both algorithms on the same 34,421 elements.
For each element $e \in D$, \erratum and WATER returned $e'_{S}$ and $e'_{W} \in D' \cup \emptyset$, respectively.
In 49.0\% of cases, \erratum and WATER agreed on a matching element ($e'_{S} = e'_{W} \ne \emptyset$).
In 13.6\% of cases, no solution found a matching element ($e'_{S} = e'_{W}= \emptyset$).
In 37.4\% of cases, \erratum and WATER disagreed on the matching element ($e'_{S} \ne e'_{W} \text{ and } (e'_{S}, e'_{W}) \ne (\emptyset, \emptyset)$).

A sample of $366$ matchings out of the $14,784$ disagreements where labelled by web testing experts, which corresponds to a 5\% confidence interval at 95\%.
Table~\ref{tab:labels} reports on the results of the manual labeling (for disagreements), thus assuming that both WATER and \erratum are correct whenever they agree.

\begin{table}[h]
    \caption{Confusion matrix on the \textsc{Wayback} dataset.}\label{tab:labels}
    \centering
    \resizebox{.65\linewidth}{!}{%
        \begin{tabular}{lcrrrr}
        \cline{3-5}
         & & \multicolumn{3}{|c|}{\textbf{\erratum}} & \\
        \cline{3-5}
         & & \multicolumn{1}{|c|}{\sf correct} & \multicolumn{1}{c|}{\sf mismatch} & \multicolumn{1}{c|}{\sf no-match} & \\
        \cline{1-5}
        \multicolumn{1}{|l|}{} & \multicolumn{1}{c|}{\sf correct}  & \multicolumn{1}{r|}{49.0\%} & \multicolumn{1}{r|}{1.5\%} & \multicolumn{1}{r|}{1.4\%} & \color{red}{51.9\%} \\
        \cline{2-5}
        \multicolumn{1}{|l|}{} & \multicolumn{1}{c|}{\sf mismatch} & \multicolumn{1}{r|}{26.5\%} & \multicolumn{1}{r|}{5.5\%} & \multicolumn{1}{r|}{3.3\%} & 35.3\% \\
        \cline{2-5}
        \multicolumn{1}{|l|}{\multirow{-3}{*}{\rotatebox[origin=c]{90}{\scriptsize \textbf{WATER}}}} & \multicolumn{1}{c|}{\sf no-match} & \multicolumn{1}{r|}{2.8\%} & \multicolumn{1}{r|}{1.9\%} & \multicolumn{1}{r|}{8.1\%} & 12.8\% \\
        \cline{1-5}
        \cline{1-5}
         & & \color{editorGreen}{78.3\%} & 8.9\% & 12.8\% & 
        \end{tabular}%
    }
\end{table}

\begin{figure}[]
  \centering
  \includegraphics[width=.75\linewidth]{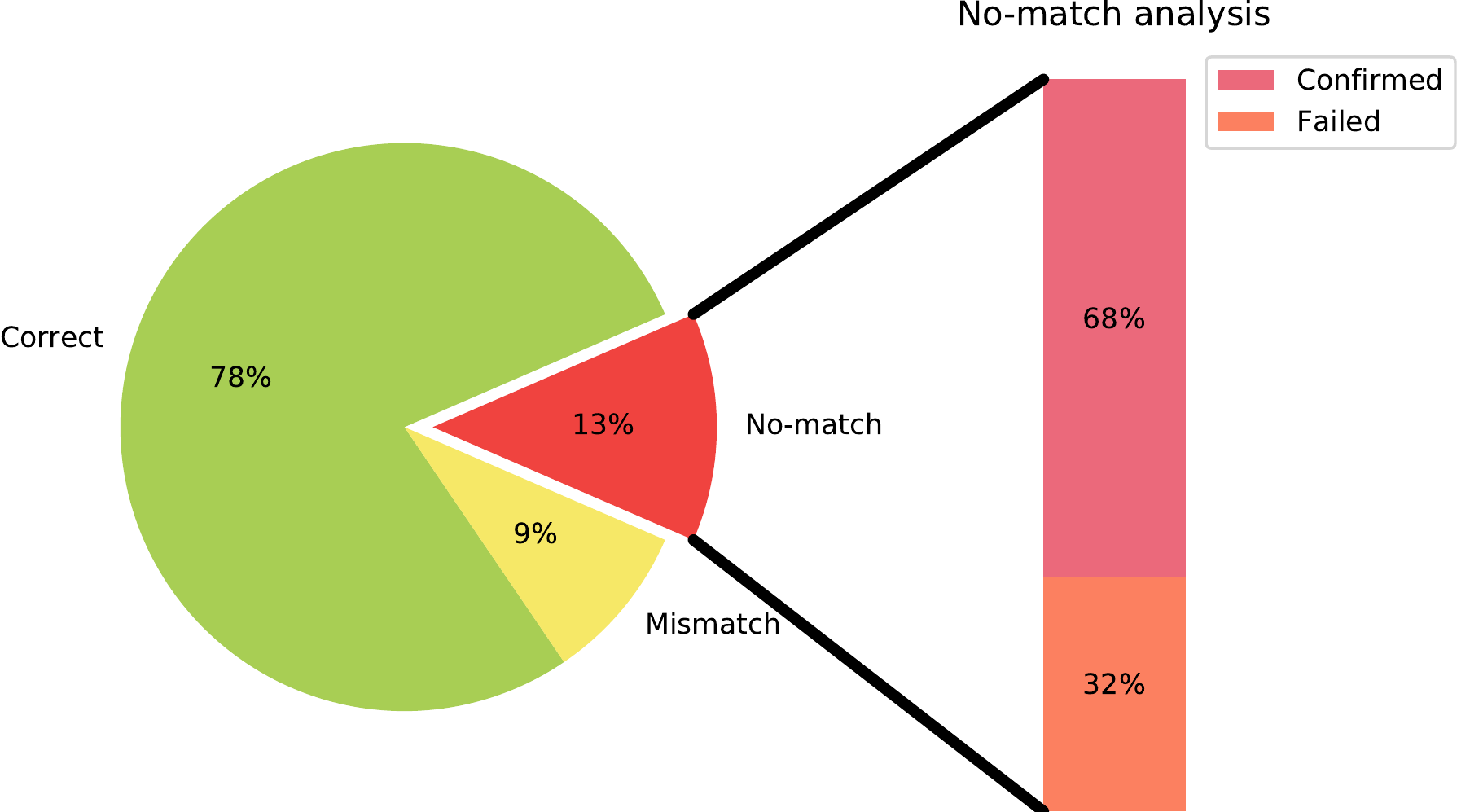}
  \caption{Analysis of matches labeled as \textsf{no-match} by \erratum{}.}
  \label{fig:nomatch-analysis}
\end{figure}

We further investigated the causes of \textsf{no-match} cases reported by \erratum{} to assess if these specific cases could be matched by experts.
As part of the \textsc{Wayback} experiment, we thus included \erratum's \textsf{no-match} cases in our labelling application (cf. Figure~\ref{fig:disagreement}) and requested the participants to eventually propose a matching element if a \textsf{no-match} was reported by \erratum{}.
The result of this evaluation, summarized in Figure~\ref{fig:nomatch-analysis}, highlights that a majority of \textsf{no-match} are accurately labeled as such by \erratum{}, since the participants could not propose a matching element in the target web page. 
For the few cases where the participants proposed a matching element, we observed that the structure of the DOM tree where subject to strong changes, thus misleading \erratum{} as already observed in Figure~\ref{fig:mutationTypeAnalysis}.

\vspace{6pt}\noindent{\bf Comparison of repair accuracy.}
Interestingly, as shown in Table~\ref{table:datasets_comparison}, the accuracy of \erratum on the {\sc Wayback} dataset ($78.3\,\% \pm 5\,\%$) is $8.7\%$ inferior to the accuracy obtained on the {\sc Mutation} dataset ($87.0\%$), while the accuracy of the WATER algorithm is better on the {\sc Wayback} dataset ($51.9\,\% \pm 5\,\%$) than on the {\sc Mutation} dataset ($42.1\,\%$) by $9.8\,\%$.
We believe the difference observed between the two datasets is because real-life mutations might not be uniformly distributed.
In particular, regarding our sensitivity analysis with regards to types of tree mutations (cf. Figure~\ref{fig:mutationTypeAnalysis}), one can guess that real-life websites are more subject to \emph{content} and \emph{attribute}-related mutations than \emph{structure}-based mutations (cf. Table~\ref{table:mutations_used}), as the former do not affect the accuracy of \erratum{}.
However, since we miss the ground truth for the {\sc Wayback} dataset, we cannot assess this hypothesis and the distribution of real-life mutations.

\begin{table}[h]
    \caption{Accuracy summary across datasets.}
    \label{table:datasets_comparison}
    \centering
    \resizebox{.5\linewidth}{!}{%
        \begin{tabular}{|l|c|c|}
        \cline{2-3}
        \multicolumn{1}{l|}{ } & {\sc Mutation} & {\sc Wayback} \\
        \cline{2-3}
        \hline
        \bf \erratum & 87.0\% & {78.3 $\pm$ 5\%} \\
        \hline
        \bf WATER    & 42.1\% & {51.9 $\pm$ 5\%} \\
        \hline
        \end{tabular}
    }
\end{table}

\subsection{Repair Time Evaluation}\label{sec:computation_time}
\begin{figure}[]
  \centering
  \includegraphics[width=.8\linewidth]{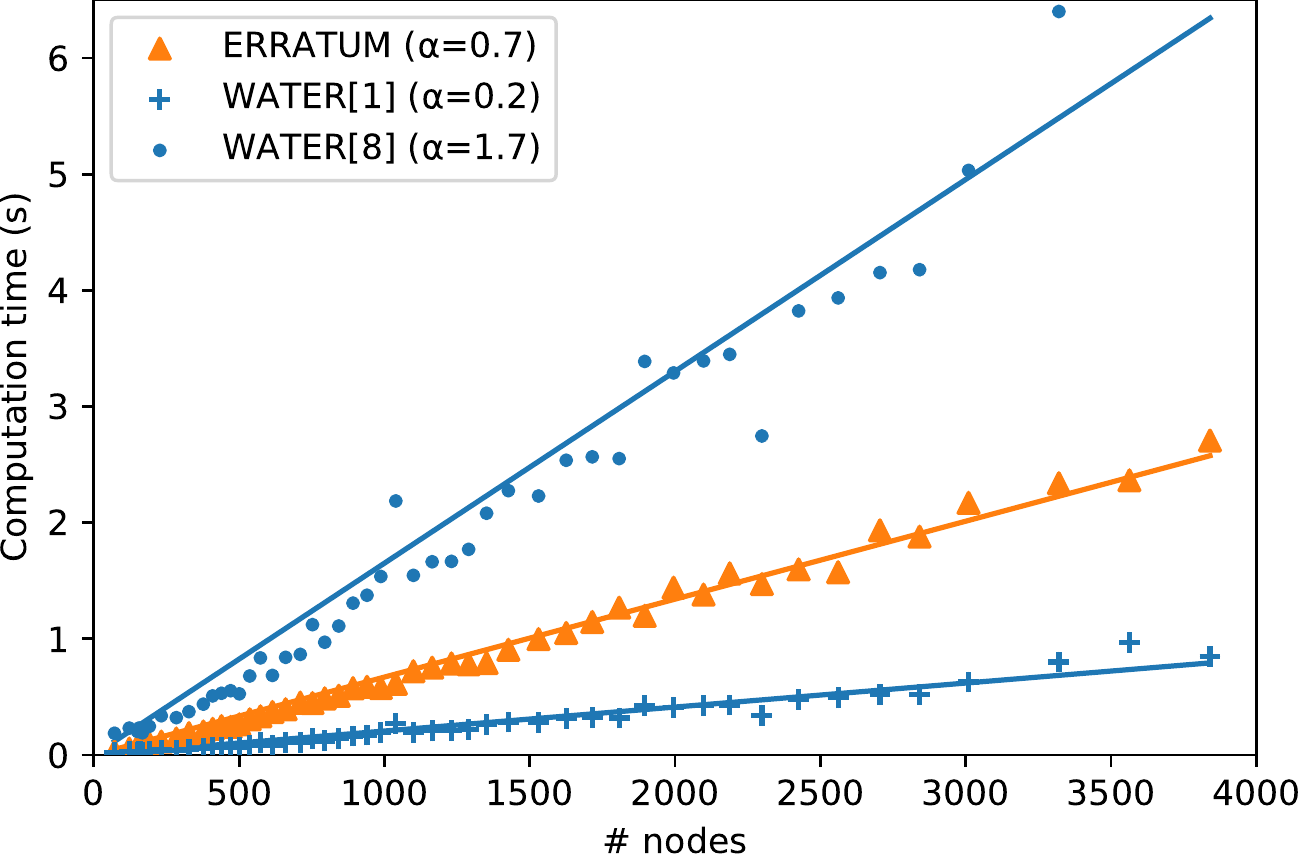}
  \caption{Repair time evolution according to DOM size.}
  \label{fig:computation}
\end{figure}

In Figure~\ref{fig:computation}, we compare the computation times of \erratum and WATER.
The results have been obtained by running both algorithms on the same server containing 252\,GB of RAM and an Intel(R)~Xeon(R)~CPU E5-2660\,v3 @ 2.60\,GHz.

ERRATUM works differently than WATER: while WATER matches one single element at a time, ERRATUM matches all elements at once. One can observe that WATER is thus faster at locating a single element than \erratum is at locating all elements.
However, when the number of locators to repair grows, the computation time of WATER evolves proportionally while the computation time of \erratum remains the same.

\begin{figure}[]
  \centering
  \includegraphics[width=.75\linewidth]{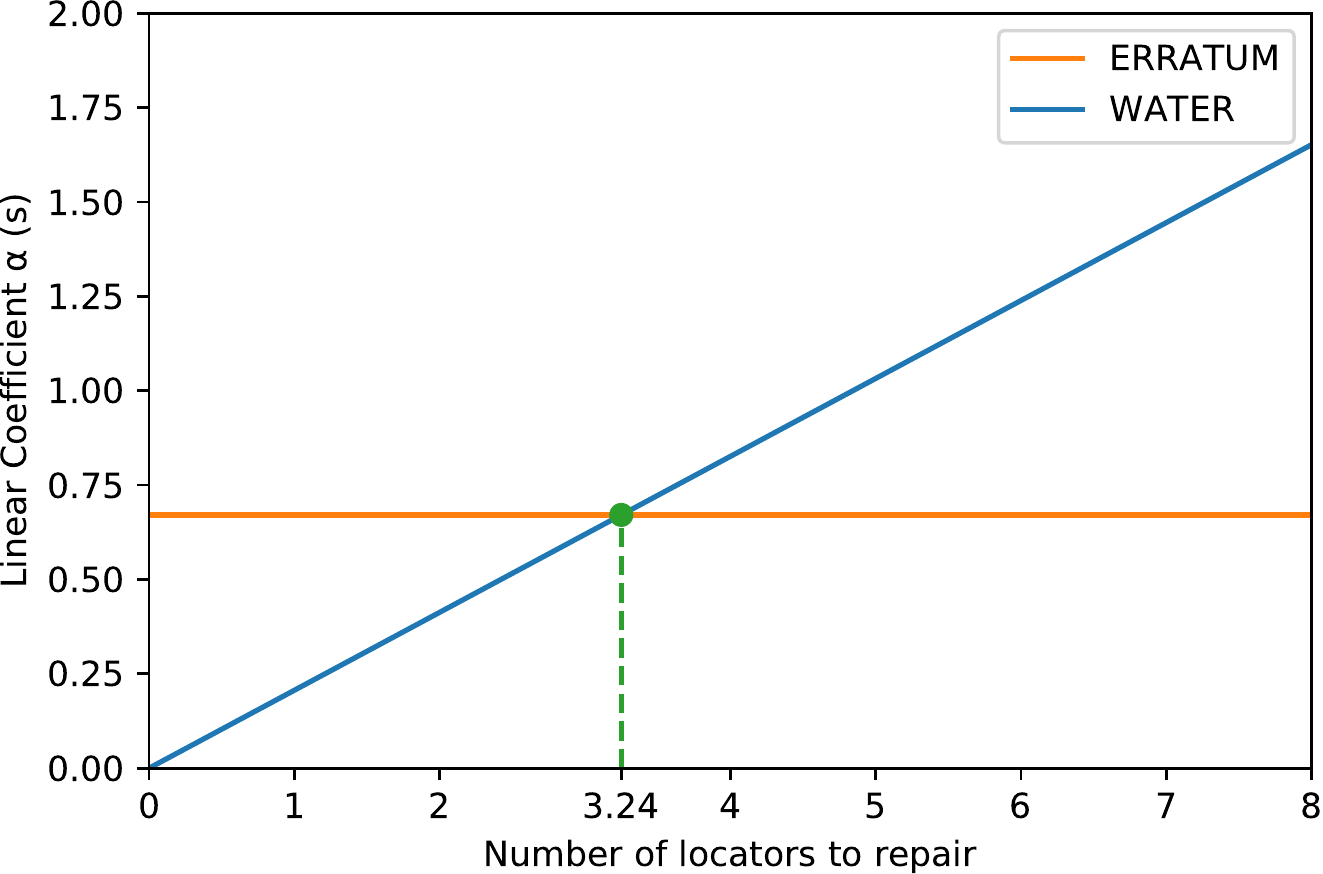}
  \caption{Performances of \erratum and WATER.}
  \label{fig:comparisonLinearCoefficients}
\end{figure}

More specifically, we compare the evolution of the performance coefficient ($\alpha$) when increasing the number of locators to repair in a web page.
Figure~\ref{fig:comparisonLinearCoefficients} plots these coefficients for \erratum and WATER, so that we can establish that \erratum{} becomes more efficient than WATER as soon as there are more than 3 locators to repair in a web page.

\subsection{Threats to Validity}\label{sec:threats}
As described in Section~\ref{sec:manualLabeling}, the \textsc{Wayback} dataset does not include a ground truth (perfect matching).
This is why we had to manually label a representative sample of the matchings obtained on this dataset, which might have introduced some bias.
To mitigate this bias, we recommended systematic and consistent decisions to label the data (cf. Section~\ref{sec:manualLabeling}). 

All our experiments with \erratum adopt the default FTM parameters, as recommended by~\cite{Kumar2011_FTM}.
Nonetheless, a thorough parameter sensitivity study would probably result in further improving the accuracy of \erratum.
Given the results we obtain on a wide diversity of web pages evolutions, we believe that this parameter tuning would only positively and marginally impact the accuracy of \erratum.

In terms of repair time, we discussed the absolute value of repair time for both solutions.
However, these values highly depend on the way each tool was implemented and the machines on which the simulations were run.
To limit this bias, both solutions were executed on the same Node.js runtime version deployed in the same environment to ensure a proper comparison.

\section{Applying \erratum}\label{sec:perspectives}
We have studied how \erratum{} can help in solving the existing locator repair problem, which is a common problem in web automation scripts.
In this section, we envision a more interactive development process made possible by \erratum{}.

When developing a web automation script, a developer typically opens the page under test in the browser, visually locates the element to interact with and then encodes (or generates) a locator for this element.
The locator is then used in the web automation script to select the target element and interact with it.
Based on the results achieved by \erratum{}, the perspectives for this work include a new layer of abstraction to the target selection.
In this new layer, web automation scripts no longer need to explicitly locate elements on the page directly, but only locate elements using a back-end service $H$:
\begin{compactenum}
    \item each web page $D$ under automation is registered in $H$,
    \item for each registered web page $D$, $H$ exposes a visual interface allowing the developer to visually select an element $e$ and retrieve its unique identifier $e_{id}$,
    \item in the web automation script, instead of using a manually encoded (or generated) XPath or CSS locator to select the target element $e$, the developer sends $(D, e_{id})$ to $H$'s API, which returns an absolute XPath selecting the target element $e$,
    \item when a web page $D$ registered in $H$ evolves into a new version $D'$, \erratum{} is automatically used to relocate all registered elements $e_{id}$ in $D$ with their matching elements in $D'$,
    \item whenever \erratum{} fails (or lacks confidence) to relocate an element, the developer is notified and invited to visually relocate the broken locator.
\end{compactenum}

This approach differs from the test repair approach described in the original WATER article. 
In the test repair approach, the locator repair is triggered by the failure of one of the test scripts.
Once such a test script fails, the test repair solution attempts to determine the cause of the breakage and if it is a locator, repair the locator.
The approach we suggest in this section does not include the analysis of any automation script, as locators are updated whenever the page changes.

In many cases, the locator breakage occurs silently (the locator is mismatched and the consequences happen only later in the test script)~\cite{stocco2018visual}.
In these situations, it is harder to locate the origin of the breakage from the test script. 
In addition to the obvious gain in time that having a visual-based breakage and repair solution provides, the \erratum-based notification process described above would help to detect such possible breakage before the scripts are even run, thus diminishing the chances for a "silent breakage" to occur.

\section{Conclusion}\label{sec:conclusion}
In this article, we considered the situation in which the evolution of a web page causes one of its associated automation scripts to break.
In the domain of automated web testing, this situation accounts for 74\% of test breakages, according to past studies~\cite{hammoudi2016record}.
Our analysis of the state-of-the-art approaches on this topic contributed to formalize the key steps involved in preventing or fixing such a kind of test breakage.

While existing solutions to the locator repair problem treats broken locators individually, we rather propose to apply a holistic approach to the problem, by leveraging an efficient tree matching algorithm.
This tree matching approach thus allows \erratum{}, our solution to repair all broken locators by mapping all the elements contained in an original page, to accurately relocate each of them in its new version at once.
To assess \erratum{}, we created and shared the first reproducible, large-scale datasets of web page locators, combining synthetic and real instances,\footnote{Dataset available from \url{https://zenodo.org/record/3800130\#.XrQb02gzY20}} which has been incorporated in a comprehensive benchmark of \erratum{} and WATER, a state-of-the-art competitor.\footnote{Benchmark available from \url{https://zenodo.org/record/3817617\#.XrWdqGgzaoQ}}
Our in-depth evaluation highlights that \erratum{} outperforms WATER, both in accuracy---by fixing twice more broken than WATER---and performances---by providing faster computation time than WATER when repairing more than 3 locators in a web test script.

\bibliographystyle{wileyj}
\bibliography{references}
\end{document}